\documentclass[a4paper,twocolumn,
secnumarabic, aps, superscriptaddress,
amssymb, amsmath, nobibnotes, showpacs]{revtex4}

\usepackage{graphicx}
\usepackage{dcolumn}

\begin{document}
\title{Modeling diffusional transport in the interphase cell nucleus}
\author{Annika Wedemeier}
\affiliation{Deutsches Krebsforschungszentrum, D-69120 Heidelberg,
Germany}
\author{Holger Merlitz}
\affiliation{Softmatter Lab, Department of Physics and ITPA, Xiamen University,
Xiamen 361005, P.R.\ China}
\author{Chen-Xu Wu}
\affiliation{Softmatter Lab, Department of Physics and ITPA, Xiamen University,
Xiamen 361005, P.R.\ China}
\author{J\"org Langowski}
\affiliation{Deutsches Krebsforschungszentrum, D-69120 Heidelberg,
Germany}

\date{\today}

\begin{abstract}
In this paper a lattice model for diffusional transport of particles in the 
interphase cell nucleus is proposed. Dense networks of 
chromatin fibers are created by three different methods:
randomly distributed, non-interconnected obstacles,   a random
walk chain model, and a self avoiding random walk chain model
with persistence length. By
comparing  a discrete and a continuous version of the random walk 
chain model, we demonstrate that lattice discretization does
not alter particle diffusion. 
The influence of the 3D geometry of the fiber network on the
particle diffusion is investigated in detail, while varying 
occupation volume, chain length, persistence length and
walker size.
It is shown that adjacency of the monomers, the excluded volume 
effect incorporated in the  self avoiding random walk model,
and, to a lesser extent, the persistence length, affect particle 
diffusion. It is demonstrated how the introduction of the
effective chain occupancy, which is a convolution of the
geometric chain volume with the walker size, eliminates
the conformational effects of the network on the diffusion,
i.e., when plotting the diffusion coefficient as a function
of the effective chain volume, the data fall onto a master curve.   
\end{abstract}


\maketitle
\clearpage
\section{Introduction}
In recent years, great progress has been made
in the view of the living cell as a regulatory
network in time. On the other hand, the coupling
of biochemical processes with the spatial 
arrangements of the cellular components is less 
understood. For a quantitative understanding of 
the function of the cell, one needs to link the
existing one dimensional genomic, functional and
metabolic information to its three dimensional 
organization. The main aspect is the quantitative
description of the transport of biomolecules 
in the dense network of macromolecules that 
constitutes the major part of the cytoplasm and
the cell nucleus.\\
Diffusive processes in the cell play a central 
role in keeping the organism alive 
\cite{organism1,organism2}. Molecules transported
through
cell membranes, drugs on their way to their protein
receptors and proteins interacting with specific
DNA sequences constituting all of the biological 
functions of DNA  \cite{alldiff1,alldiff2}
 are diffusion controlled reactions. Furthermore, 
proteins approaching their specific 
target sites on DNA are transported by diffusion
or even facilitated diffusion \cite{Holger1, Holger2,fd1, fd2, fd3, fd4}.
For the 1D sliding motion experimental evidence
exists \cite{exp1, exp2}.\\ 
However, diffusional transport of molecules in the 
living cell is fundamentally different from the 
normal kind of diffusion which a molecule undergoes
in a homogeneous fluid where the mean square 
displacement of a molecule behaves linear in time $t$, 
$\langle r^2(t) \rangle =6Dt$ with $D$ as  the 
diffusion coefficient. The cellular environment is a 
crowded solution of macromolecules. In particular, the
interphase cell nucleus constitutes a dense network of
chromatin fibers with a volume fraction ranging from 
5\% to 12\%. The motion of other macromolecules is 
strongly influenced by the presence of this "sticky
tangle" due to steric obstruction and transient 
binding. Fluorescence 
correlation spectroscopy (FCS) studies have shown 
obstructed diffusion of autofluorescent proteins \cite{wachsmuth, banks}.
Subdiffusion was also reported for other systems,
for instance mRNA molecules and dense dextrans \cite{examp1, examp2}.
Obstructed diffusion or
subdiffusion is characterized by 
$\langle r^2(t) \rangle \sim t^{\alpha}$ with
the anomalous diffusion exponent $\alpha <1$. 
Subdiffusion like that might give rise to a weak ergocidity
breaking, causing interesting effects in intracellular diffusion
of macromolecules \cite{metzler}. \\
 Other FCS measurements indicate that most 
of the nuclear space is accessible to medium sized 
proteins by simple diffusion and that there is no 
preference for interchromosome territory channels \cite{weidemann}.
In general, macromolecular transport in the living cell
nucleus is only rudimentarily understood. 
In particular, it is still a matter of intensive
discussion \cite{wachsmuth}
 to what extent
 macromolecular mobility is affected by structural 
 components of the nucleus.\\
 This paper shall contribute to the understanding of this 
process by developing a theoretical description of 
network diffusion in the interphase cell nucleus. 
There exist already theoretical descriptions of diffusion
in the cell nucleus \cite{subdiffusion}.
 However,
these approaches do not incorporate realistic structures 
of the chromatin fibers. In the following the chromatin fibers are
referred to as (polymer) chains and proteins diffusing through
the fiber network are referred to as particles or walkers.
In the present paper we 
investigate in detail how the diffusion coefficient of
diffusionally transported particles of various size depends
on the 3D geometry of the network
of chromatin fibers and their density  in the nucleus. 
Furthermore, we investigate to what extent structural 
properties of the fibers such as persistence length and
contour length influence the diffusion coefficient.\\
The chromatin network in the cell nucleus during
interphase is modeled using a lattice approach minimizing  
computational time and effort. The first method creates a crowded
environment  with polymer chains constructed by a random walk (RW)
on the lattice without excluded volume. 
To confirm the accuracy of the lattice model, 
these results are also compared to simulations of the
corresponding continuum model. Later, we introduce a self avoiding
random walk (SAW) of well equilibrated 
polymer chains with excluded volume interaction, 
which deliver more realistic static properties such as 
end-to-end distance. 
These chains are simulated on the lattice by applying a simplified 
version of the bond fluctuation method  \cite{BFM},
  the single 
site model, in combination with a Metropolis Monte Carlo (MC) procedure \cite{MC}. 
The bond fluctuation model (BFM) was  introduced 
as a new effective algorithm for the dynamics of polymers 
by Carmesin et al. \cite{BFM} 
and provides a very effective 
means to simulate the dynamic behaviour of e.g.\ dense
polymer melts \cite{Sommer1,Sommer2}.
Both methods, with RW and SAW chain model, 
are compared to a third test system
consisting of disconnected, randomly distributed obstacles.\\
In section 2 the lattice for chain construction and particle
diffusion is presented. A brief overview of the modeling of
particles of different sizes is given, reflecting the diffusing 
proteins in the cell nucleus. In addition, the test system
is described.\\
Section 3 introduces the RW chain model for the discrete
and continuous space. We test the validity of the lattice
approach by comparing the diffusion coefficient of particles in
the discrete and the continuous model.
Afterwards, the chain relaxation simulation is described. 
After presenting  the results on particle diffusion
in the RW chain model and the test system we conclude 
that the latter system is 
insufficient for a description of a chromatin fiber network.  
In section 4 the SAW chain model is introduced exhibiting
important features of a real biological fiber network such
as persistence length. The chain simulation algorithm is
presented explaining shortly the BFM and MC procedure.
It is shown that the static properties of the 
polymer chains in the SAW model  agree with known scaling laws.
Anomalous translational diffusion of the chains' center of mass
is observed. 
 After stating the results on particle diffusion and comparing
them to the RW chain model, it is concluded that the SAW
chain model comes closer to the situation in the cell nucleus
and thus yields more realistic diffusion coefficients. 
Finally, it is  found that the diffusion coefficient of the particles 
depends on the persistence length of the fibers, but not on their 
contour length, as long as they
consist of connected structures of several monomers. 
However, it does depend on the particular
geometry composed by well equilibrated chains created by
the SAW chain model. Section 5 summarizes the obtained results.

\section{Modeling volume}
The model is contained in a $100\times100\times100$  cubic
lattice. The  penetrability of the lattice walls is different regarding
 chains and particle motion and is discussed in the following.

\subsection{Particles}
 Proteins transported in the nucleus differ, among other things, in 
their size. 
For instance tracer particles with a size of 
 27 kD to 282 kD were used to study diffusional transport \cite{banks}.
  In our model systems
 particles of different sizes are implemented by a
corresponding numbers of occupied  lattice sites in a cubic 
arrangement. We use three particle sizes: one occupied lattice
site (small), a $2^3$ cube (medium) and a $3^3$ cube (large).
In the continuous model spherical walkers of three different 
sizes are tested, the smallest walker being point-like with a 
radius $R_{walker}=0$, the medium sized particle with a radius $R_{walker}=1$ 
and the largest walker with a radius $R_{walker}=2$. 

\subsection{Diffusional transport simulation}
The movement of a particle is modelled by a random walk. 
Particles are allowed to visit only unoccupied lattice sites. 
If a particle collides with a chain it is reflected to the last 
visited lattice site. The initial position is sampled randomly
among non-occupied lattice sites. Then, random walks of different
duration are carried out. To prevent boundary effects due to the 
lattice walls, periodic boundary conditions are applied.

\subsection{Test system with disconnected obstacles}
 To study the influence of chain connectivity  on particle diffusion,
 a test system is introduced with randomly
distributed obstacles, occupying between 5\% and 15\% of
the lattice sites.

\section{Random Walk Chain Systems}
In this system the chromatin fibers are created by a standard random
walk without excluded volume effects, i.e.\ allowing self intersection.
This approach is more realistic than the set of disconnected obstacles
of the test system, since here the monomers are interconnected to
create fibers. This system is set up in both, lattice and continuum
models, in order to justify the accuracy of the lattice approximation.

\subsection{Model and chain simulation}
In the discrete model the chromatin fibers are realized by chains of 
occupied connected lattice sites chosen by random walks on the 
$100\times 100\times 100$ lattice. 
In the continuous model the modeling volume is defined by a sphere
of radius  $R_{sphere}=50$ containing a chain of given 
length implemented by a random walk. 
To simulate a chain in the lattice model one site is chosen randomly 
as a starting 
point of the chain. During each random walk step one of the 6 
nearest neighboring lattice sites is randomly selected with equal 
probability and is occupied. The walls of the modeling volume are reflecting, 
no flux boundary conditions are applied. The total number of occupied 
lattice sites yields the occupation volume. Simulations 
of random walks with lengths between 100000 and 700000 steps
were carried out  yielding occupation volumes between 6\% and 36\%
(effective occupation volumes between 18\% and 78\%). The effective
 volume is defined in the next subsection.\\
 In the continuous model a chain is created with a radius  $R_{chain}=1$ 
 using a random walk with step size of 5, which was chosen due to lower
 computational cost instead of a smaller step size. This is justified because
 of the self-similarity of the random walk which implies that the step size
 is irrelevant if the modeling volume is sufficiently large. 
 The step size equals the segment length.\\
 The volume exclusion effects are implemented in the following way:
The chain is made of segments of radius $r$. The total geometric
volume of the chain is computed via Monte Carlo integration
(throw $10^9$ points into the system and count how many of them
are inside the chain).

The walker is a sphere of radius $R$. Collision with the chain
takes place if the distance to the chain is $< R+r$, but this
is equivalent to a chain of thickness $R+r$ and point-like
walker (because we use a simple random walk for the chain
contour and do not consider chain-chain excluded volume). So
the effective volume is again found via MC integration, but
this time with chain radius $R+r$.

 \begin{figure}[h]
\centerline{
{\bf a }
\includegraphics[scale=.3]{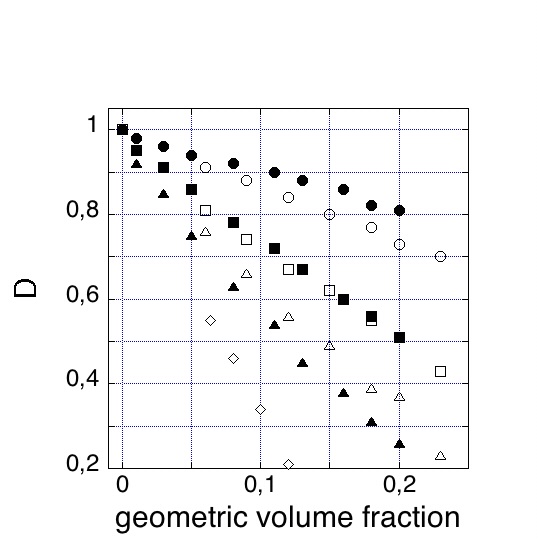}
}
\centerline{
{\bf b}
\includegraphics[scale=.3]{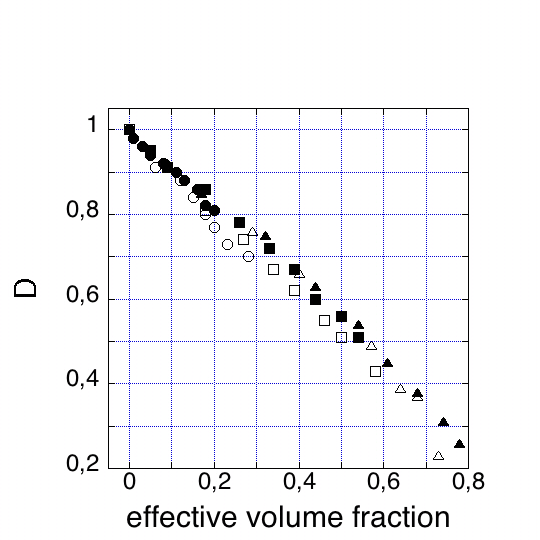}
}
\caption{
 Comparison of the continuous and discrete 
random walk chain model and the test system (disconnected obstacles).
Upper panel: Diffusion coefficient dependent on the geometric
volume fraction.
Lower panel: Diffusion coefficient dependent on the effective 
volume fraction.
Circles: smallest
particle, squares: medium sized particle, triangles: largest
particle, and diamonds: medium sized particle in the test system.   
Blank: Discrete random walk chain model.
Solid: Continuous random walk chain model.
\label{fig: effvolumecondis}}
\end{figure} 
\subsection{Results}
\begin{figure}[h]
\centerline{
\includegraphics[scale=.3]{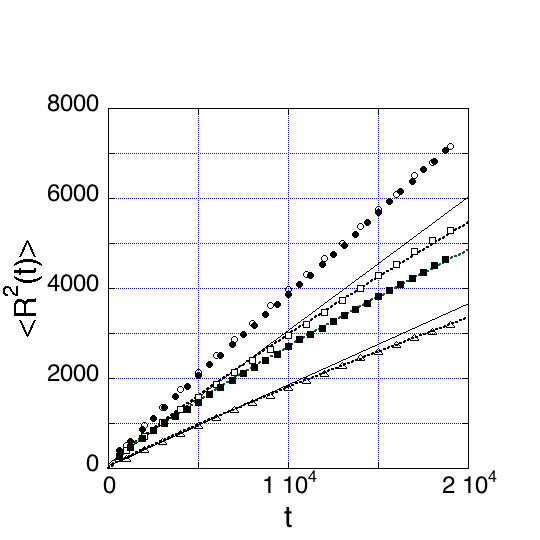}
}
\caption{Mean square displacement of the largest particle vs. time.
The continuous lines are linear fits through the first five points of the curve
(blank squares, blank triangles) to illustrate the deviation
from linearity for the long-time diffusion in the denser systems. The dashed lines
are power law fits yielding the anomalous diffusion exponent $\alpha$.
Blank circles: Random walk chain model (effective volume 68\%, geometric
volume 20\%).
Solid circles: Continuous random walk chain model  (effective volume 68\%, geometric
volume 20\%).
Blank squares: Random walk chain model (effective volume 73\%, geometric
volume 23\%, $\alpha=0.88$).
Solid squares: Continuous random walk chain model  (effective volume 78\%, geometric
volume 26\%, $\alpha=0.85$).
Triangles: Self avoiding random walk chain model (effective volume 81\%, 
geometric volume: 12.5\%, chain length $N=50$, $l_p=0.5$, $\alpha=0.89$).
\label{fig: msd}}
\end{figure}

In the following the occupation volume of the lattice is either computed 
as the geometric volume or the effective volume. The geometric volume
 is defined as the total number of  occupied lattice sites. The effective 
 volume is defined as the space which is not freely accessible to a particle
  of given size. Hence, the effective volume of the chains on the lattice 
  depends on the particle size. For a particle consisting of one
occupied lattice site the geometric volume of the chain equals its effective 
volume.\\
One particle was randomly initialized in the lattice and in the spherical 
volume. 200000 random walk steps were carried out and the motion of
the walker was monitored. This procedure 
was repeated 5000 times and the data were averaged. 
Fig. \ref{fig: effvolumecondis}  shows the dependence of the diffusion 
coefficient on the geometric volume 
fraction and on the size of the particle in the lattice model and in the 
continuous model. In both models  subdiffusion 
was observed around a geometric volume of 23\% for the largest 
walker (Fig. \ref{fig: msd}), indicating it to be occasionally trapped. 
The anomalous diffusion exponent $\alpha=0.88$ was found by
fitting the curve "mean square displacement vs. $t$" with a power law.
The non-adjacency of 
the occupied lattice sites in the test system was made responsible for a 
decreased diffusion coefficient compared to that of the random walk chain
model: At a given value of the occupied geometric volume, the obstacles
of the test system are more homogeneously distributed than the RW chains. This
property is yielding a larger value for the effective occupation volume,
so that walkers of large size are unable to find a percolation path around
the obstacles. To the contrast, the random walk chains create regions
of high density which are not accessible to the walker, but also leave 
areas of low density where a diffusion remains fairly unaffected. \\
Fig. \ref{fig: effvolumecondis}  also shows the linear dependence of the diffusion 
coefficient on the effective volume fraction and on the size of the particle
in the lattice model and in the continuous model respectively. The values lie
on a straight line besides some slight deviations in the discrete model 
which may be due to the lattice discretization. This result reveals that in both 
models the effect of increasing either the size of the particle or the occupation 
volume of the chain is the same. Moreover,
\begin{equation}
D\propto\sigma_{free},
\end{equation}
where $\sigma_{free}$ describes the volume fraction of the freely accessible 
space for a particle of given size.\\ 
The discrete and continuous model agree very well with another. 
The dependence of the diffusion coefficient on the occupation
volume and the size of the walker is quantitatively the same in both models.
Moreover, in both models subdiffusion is found in the same 
ranges of the geometric and effective occupation volume (Fig. \ref{fig: msd}).

\subsection{Conclusions: RW chains}
Due to the very good agreement of the results obtained with the
discrete and the continuous model,  we conclude that
lattice discretization does not affect the characteristic
properties of particle diffusion which are of interest in this
work. Thus, for the rest of our investigations, the lattice model,
being between one and two orders of magnitude faster than a corresponding 
continuum model, is employed. However, the 
adjacency of the chain monomers has an impact on particle diffusion. Hence, 
the test system, consisting of randomly distributed occupied lattice sites, 
is not a reasonable approximation to the properties of a 
crowded environment inside the cell nucleus.

\section{Self Avoiding Random Walk Chain System}
The SAW chain system incorporates two important characteristics
of chromatin fibers in a biological system which are not considered
in the RW chain model.\\
In the SAW chain system, chains are created by a self avoiding random
walk, i.e. the excluded volume effect is taken into consideration.
The chains are well equilibrated, relaxed and satisfy more realistic
static properties such as end-to-end distance.
Moreover, the chains exhibit a given persistence length reflecting a 
certain stiffness of the fiber. \\
Comparing the results on particle diffusion in this system to those of the
RW chain system shows the dependence of the particle diffusion
on the excluded volume effect and the persistence length.

\begin{figure}[t]
\centerline{
{\bf a }
\includegraphics[scale=.3]{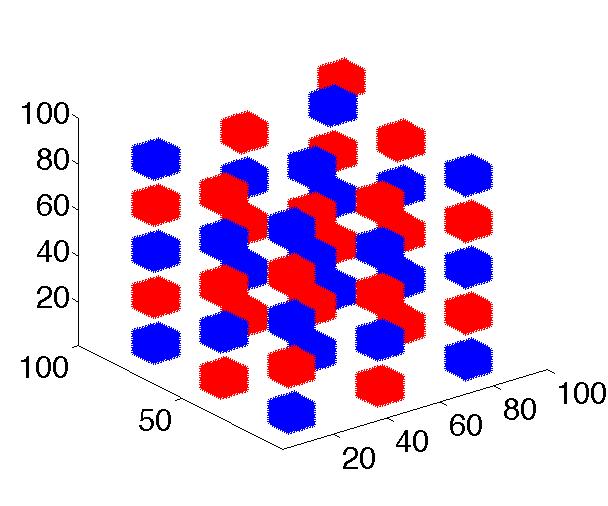}
}
\centerline{
{\bf b}
\includegraphics[scale=.3]{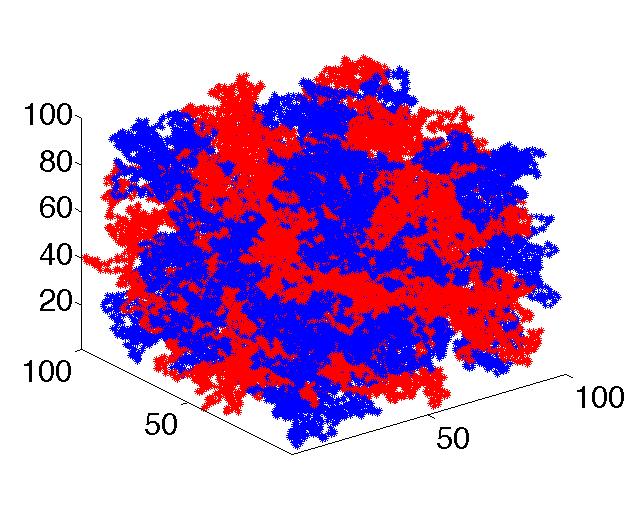}
}
\caption{Upper panel: Start conformation of the Monte Carlo algorithm. 
46 cubes, chains with $N=12^3$, are homogeneously distributed in the 
universe. Lower panel: Relaxed conformation after $ 6\cdot10^{9}$ time steps
of the Monte Carlo algorithm combined with the bond fluctuation method.
The chains constitute a dense network, comparable with chromosome
territories inside the cell nucleus.  To highlight the topological structure
of the chain network, the 46 chromosomes are alternately coloured,
here red and blue.
\label{fig: conformation}}
\end{figure} 
\subsection{Model}
 The chromatin fibers are
modelled by chains consisting of monomers connected by segments.  
Monomers are represented by occupied lattice sites. Neighboring 
occupied lattice sites reflect neighboring monomers in the chains. 
No lattice site is allowed to be occupied more than once.
Possible bonds between two adjacent monomers are given by the 
set of all component permutations and sign inversions $P$ of two 
bond vectors:
$$P(1,0,0)\cup P(1,1,0).$$
The initial state for the Monte Carlo process consists of cubic compact
 chains. Cubes in the initial state have the dimension $L\times L\times L$ 
 and are periodically distributed in the lattice (Fig. \ref{fig: conformation}). \\
Every cube consists of $L$ planes of $L\times L$ occupied lattice sites. 
The occupied sites are connected in a way that a Hamiltonian path results. 
If $L$ is even, every plane of the cube is rotated by  $90^{\circ}$ with 
respect to the previous one. Otherwise, if $L$ is odd, every second 
plane of the cube is rotated by $90^{\circ}$  with respect to the first one. 
The Hamiltonian path is afterwards reflected at the diagonal of the plane. 
Start and end point of the path remain fixed during such a reflection. \\
The planes of one cube are connected with each other so that the chain 
of a cube represents a Hamiltonian path, a chain with $L \times L \times L=N$ 
monomers. \\
The energy model for the stiffness of the chromatin fibers includes a 
bending potential $E^{b}$. This potential is computed as
\begin{equation}
E^{b}=\gamma \sum_{i=1}^{N-2}{\theta}^2_{i}.
\end{equation}
$\theta_i$ is the angular displacement of bond $i$ relative to bond $i+1$.
$\gamma$ in units of $k_BT$ is the bending rigidity constant and is directly 
related to the persistence length $l_p$ of a fiber. $\gamma$ is determined 
by the procedure described in Jian et al. \cite{schlick}.

\subsection{Chain simulation}
To simulate equilibrium conformational distributions of chromatin fibers on
the lattice, a combined algorithm of the simplified bond fluctuation method
and the Metropolis MC procedure is applied.

\subsubsection{Bond Fluctuation Method - single site model}
We first briefly describe the simplified version of BFM, the single
site model. The fiber is moved by local jumps of the monomers.
 The number of monomers is fixed but the bond length $l$ is variable 
 up to some restrictions to avoid bond cuts. In the single site model 
 either $l=1$ or $l=\sqrt2$ are permitted. One monomer of the fiber is randomly 
 selected. It tries to jump randomly the distance of one lattice unit
  into one of the possible lattice directions. If the bond length restrictions 
 are fulfilled and the newly chosen lattice site is unoccupied 
 (self avoiding walk conditions are fulfilled), the move is accepted. 
Otherwise a new monomer is randomly selected. The boundaries of the 
lattice are reflecting.

\subsubsection{Metropolis Monte Carlo procedure}
The allowed moves of the single site model are used as perturbation
 moves. The standard rules of Metropolis et al. \cite{MC}  
 yield the 
 probability of accepting a new conformation. If the total bending potential 
 energy $E_{new}$ is lower than $E_{old}$, the energy of the previous 
 conformation, the new conformation is accepted. If $E_{new}$ is larger 
 than $E_{old}$, the probability of acceptance of the new conformation
  can be expressed as $p_{acc}=\exp(-\gamma(E_{new}-E_{old}))$. 
  The probability of acceptance $p_{acc}$ is therefore  compared to a
 random  number $a$ uniformly distributed in $[0,1]$. If $a>p_{acc}$ the 
 new conformation is accepted. Otherwise a new move according to the 
 single site model is induced and the old conformation is counted
 once more.\\
With different values of $\gamma$ used in the computations, acceptance 
rates between 10\% and 60\% of the Metropolis Monte Carlo procedure
 were observed. The acceptance rate did neither depend on the 
 investigated occupation volume of the lattice nor on the chain length 
 of the fibers. \\
The combined BFM-MC procedure applied to a typical start conformation
 yields a "sticky tangle", a dense network of chains (Fig. \ref{fig: conformation}). 

\subsection{Results}
In the next subsections it is verified that the SAW chains were well equilibrated
and relaxed. Moreover, it is shown that the translational diffusion of chains
behaves anomalous.

\subsubsection{Chain relaxation}
\begin{figure}[h]
\centerline{
\includegraphics[scale=.3]{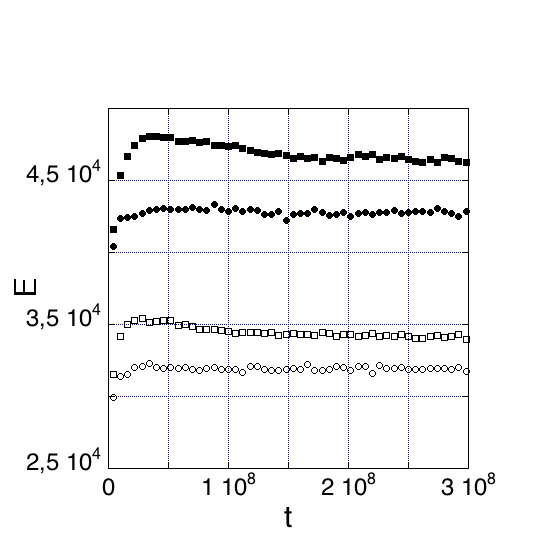}
}
\caption{ Energy distribution E as a function of time  $t $.
Geometric occupation volume: $6.4\%$.
 Circles: $N=25 $, squares: 
$N=200$. Blank: $l_p=1 $, solid: $l_p=1.5 $.
\label{fig: energy}}
\end{figure}

\begin{figure}[h]
\centerline{
\includegraphics[scale=.3]{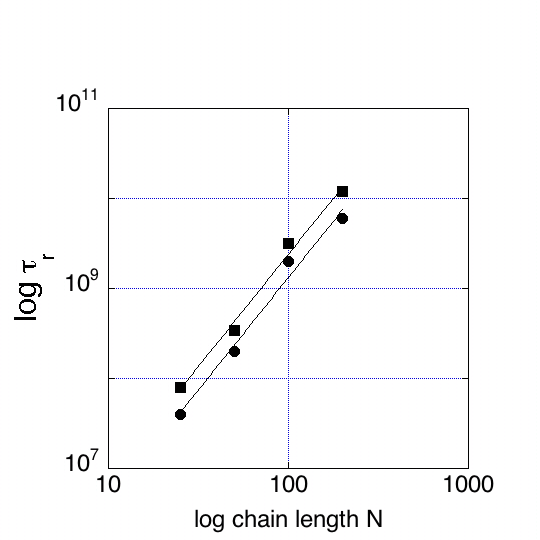}
}
\caption{Relaxation time $\tau_r$ vs chain length $N$ on a
logarithmic scale. 
Geometric occupation volume: $6.4\%$.
The solid line is a power law fit with
an exponent of 2.5. Squares: $l_p=1$,
circles: $l_p=0.5 $.
\label{fig:tau}}
\end{figure}

Using the energy of the system as an indicator for chain relaxation,
the chains seem well equilibrated after $3\cdot10^{8}$ BFM-MC steps,
independent of the persistence length $l_p$ of the chains (Fig. \ref{fig: energy}).
However, the relaxation of the average end to end distance is slower 
and depends stronger on $l_p$. It is characterized as 
$\left\langle R(t)^2\right\rangle$ which is defined as 
$\left\langle R(t)^2\right\rangle=\left\langle (r_1(t)-r_N(t))^2\right\rangle$.
$r_1(t)$ and $r_N(t)$ denote the position of the first monomer of the chain 
and the last monomer of the chain at time $t$ respectively.
Here and in the following, the brackets 
$\left\langle \right\rangle$ indicate ensemble averaging.\\
 When simulating chains of more than 1000 monomers $(N>1000)$, 
the finite size of the lattice as well as entanglement of the chains 
influence the relaxation. In this case,
\begin{equation}
\left\langle R(t)^2\right\rangle < \left\langle R^2\right\rangle_{max},
\end{equation}
with $\left\langle R^2\right\rangle_{max}\approx 1500$
was found regardless of the persistence length. Although such kind of boundary
effects may appear spurious on the first sight, they are actually not, since
inside a real cell nucleus the fibers are also confined within a finite
volume, and boundary effects on particle diffusion might become a factor
in experimental measurements as well.\\
For the equilibrium state,
\begin{equation}
\left\langle R(t)^2\right\rangle_{\infty}=l_0^2N^{2\nu},
\end{equation}
where $0.50< \nu <0.55$ was observed if
$\left\langle R(t)^2\right\rangle_{\infty} < <\left\langle R^2\right\rangle_{max}$. 
$ \l_0$ is the average bond length and $\nu$ is the universal scaling
constant. The fact that $\nu > 0.5$ is a result of the excluded volume 
effects. For a single chain, a value near Flory's $\nu \approx 0.6$ would be
expected, but for the concentrations reached here, a semi-dilute scaling
takes over, which, in the long chain limit, again approaches a value near 
$\nu = 0.5$.\\ 
The relaxation time $\tau_r$ is defined as the number of time steps needed to 
obtain relaxed fibers during simulation. Reaching relaxed fiber conformations is
characterized by a plateau of energy and of the end-to-end distance. We found that
$\tau_r$ scaled with the chain length, see Fig. \ref{fig:tau}, as
\begin{equation}
\tau_r \propto N^{2.5}.
\end{equation}

\subsubsection{Translational diffusion coefficient}
\begin{figure}[h]
\centerline{
{\bf a }
\includegraphics[scale=.3]{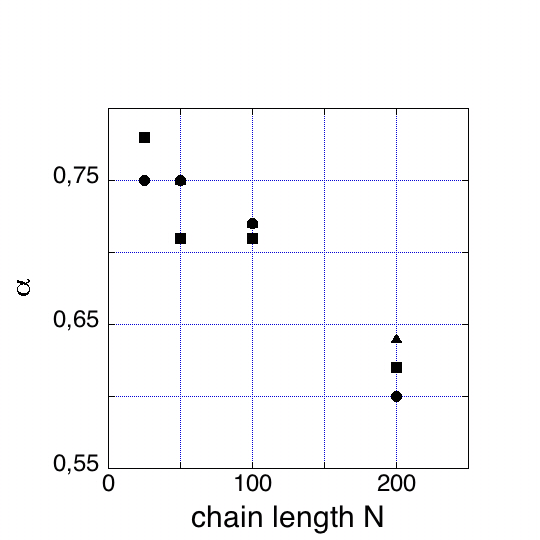}
}
\centerline{
{\bf b }
\includegraphics[scale=.3]{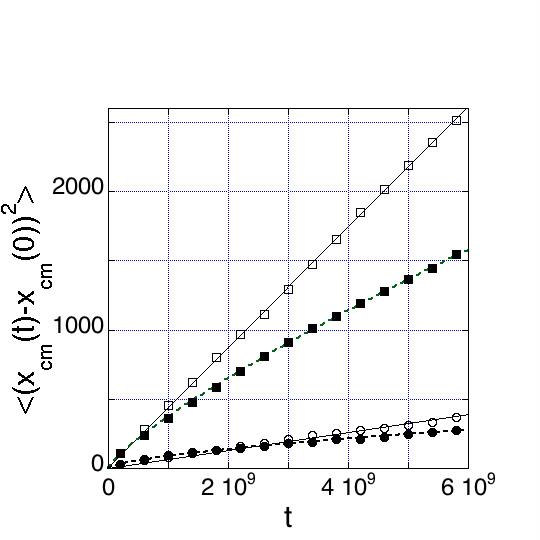}
}
\caption{Upper panel: Anomalous translational diffusion exponent $\alpha$
as a function of  chain length $N$. 
Geometric occupation volume: $6.4\%$.
Circles: $l_p=1.5 $, squares: 
$l_p=1$, triangles: $l_p=0.5$.
Lower panel: Center of mass 
$ \left\langle(x_{cm}(t)-x_{cm}(0))^2\right\rangle$ vs. time $t$. 
Squares: $N=25$, $l_p=0.5$, circles: $N=200$, $l_p=0.5$. 
Solid: $100\times100\times100$  cubic lattice, 
blank: $500\times500\times500$  cubic lattice. 
Straight line: linear fit, dashed line: power law fit.
\label{fig:alpha}}
\end{figure}

The translational diffusion coefficient $D_{trans}$ 
of the chain can be
estimated by investigating the averaged movement
of its center of mass  via the Einstein-Stokes equation
\begin{equation}
6tD_{trans}=\left\langle(x_{cm}(t)-x_{cm}(0))^2\right\rangle.
\end{equation}
$x_{cm}(t)$ is the center of mass position vector of one
chain at time $t$. 
In case of anomalous translational diffusion,
\begin{equation}
t^{\alpha} \propto \left\langle(x_{cm}(t)-x_{cm}(0))^2\right\rangle
\end{equation} holds
with
\begin{equation}
D_{trans}(t)\propto 1/t^{1-\alpha}.
\end{equation}
 A reduction of the anomalous diffusion exponent $ \alpha$  
 and therewith of the diffusion coefficient is observed
when increasing the length of the chains (Fig. \ref{fig:alpha},
upper panel). This anomalous 
diffusion is a result of the fixed boundary of the lattice.
Probing chain diffusion in a much larger volume
(a $500\times500\times500$  cubic
lattice, which can be considered as practically infinite), 
yields normal diffusion (Fig. \ref{fig:alpha}, lower panel). 
As mentioned above, such kind of boundary effects may appear spurious, 
and could technically be resolved when using periodic boundary conditions.
However, a cell nucleus does not represent a system of infinite
matter, but actually has fixed boundaries. Each chromatin fiber has
to squeeze inside a limited volume known as chromosome territory
(Fig. \ref{fig: conformation}).

Once the lattice is large enough to avoid boundary effects,
the translational diffusion $D_{trans}$  decreases roughly 
inversely proportional to the chain length
$N$, as is to be expected for simulations without
hydrodynamic interaction, yielding Rouse scaling \cite{rouse53}.
The neglect of hydrodynamic interaction is a reasonable
approximation for systems with high chain concentration.
In fact, a slowdown of diffusion with increasing chain length 
has been experimentally verified with DNA fragments in solution \cite{slowdown}.
From the experimental point of view, however, it is not clear yet 
which contribution to the slowdown came from effects other than trivial
friction, e.g. entanglement or binding activities of the chains.
Numerical models as presented here will be useful to investigate
these effects in detail, since parameters which control features 
like binding affinity can be easily modified in systematic 
simulations.
\subsubsection{Particle diffusion in various environments}
The initial position  was randomly sampled inside the lattice,
avoiding occupied lattice sites. 
Random walks of short times, 600 time steps, and long times, 
200000 time steps, were carried out. This procedure was 
repeated over 5000 times and the data were averaged.
The occupation volume was varied between 6.4\% and 12.5\%.\\

\begin{figure}[t]
\centerline{
{\bf a }
\includegraphics[scale=.3]{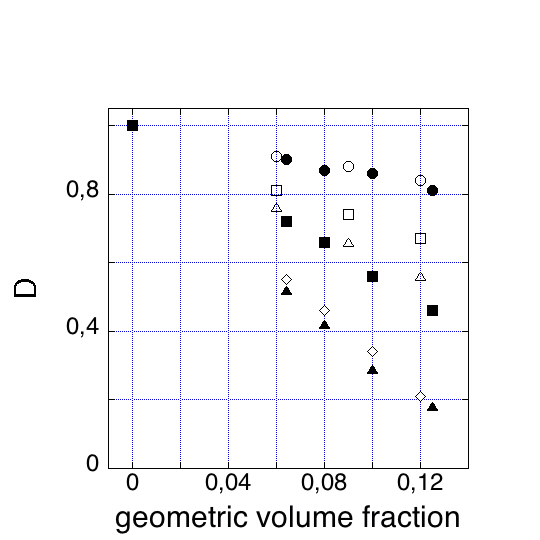}
}
\centerline{
{\bf b}
\includegraphics[scale=.3]{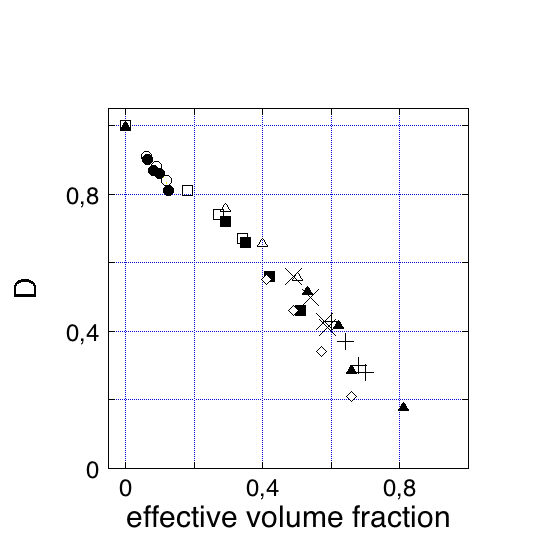}
}
\caption{Comparison of the random walk chain model and the
self avoiding random walk chain model.
Upper panel: Diffusion coefficient dependent on the geometric
volume fraction. Lower panel: Diffusion coefficient dependent on the 
effective volume fraction. Blank: Random walk chain model.
Solid: Self avoiding random walk chain model. 
 Diamonds: middle sized particle in the
test system.
 Circles: smallest
particle, squares: middle sized particle, and triangles: largest
particle. 
Stars: largest particle in the SAW system with different 
persistence lengths $l_p$ of 0.5, 1, 2 and 3 with constant
geometric occupation volume of $6.12 \%$, respectively.
Crosses: largest particle in the SAW system with different 
persistence lengths $l_p$ of 0.5, 1, 2 and 3 with constant
geometric occupation volume of $7.95 \%$, respectively.
\label{fig:geoeff}}
\end{figure} 

Figure \ref{fig:geoeff} displays the dependence of the diffusion 
coefficient on the geometric occupation volume and the walker
 size (solid symbols). Similar to the RW simulations (blank symbols) and
the test system using disconnected obstacles (diamonds),
a linear dependence of the diffusion coefficient on the effective 
occupation volume is found. Again, particles of larger size
displayed a slower diffusion (upper panel), but additionally a clear 
separation regarding the environment is visible: Using RW chains,
the particle was diffusing faster compared to SAW chains of identical
(geometric) occupation volume. Slowest diffusion was observed
using the test system of entirely disconnected obstacles (diamonds,
only data for the medium sized particle are shown). 

The reason for this separation: A SAW leads to a rather homogeneous
chain distribution, due to the excluded volume interaction, when
compared to the uncorrelated RW chain. Disconnected obstacles are
even more randomly distributed. This is affecting the distribution
of pore sizes and hence percolation paths through unoccupied lattice points.
In particular, a random walker of large size is more restricted
inside the SAW environment than inside the RW fibers, the latter
offering large density fluctuations and hence larger connected 
pores of free space.      

Again, these differences were properly accounted for after rescaling
with respect to  the effective volume fraction of the chains 
(Fig.\  \ref{fig:geoeff}, lower panel). Now, all data roughly fell 
onto a single master curve (lower panel). This result indicates that,
in fact, different chain densities and conformations affect particle
diffusion, but these subtle differences can be projected into a
single parameter, the effective volume fraction.

\begin{figure}[t]
\centerline{
{\bf a }
\includegraphics[scale=.3]{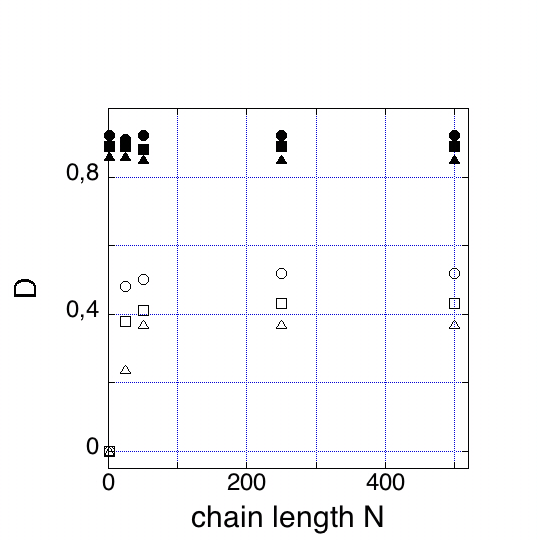}
}
\centerline{
{\bf b}
\includegraphics[scale=.3]{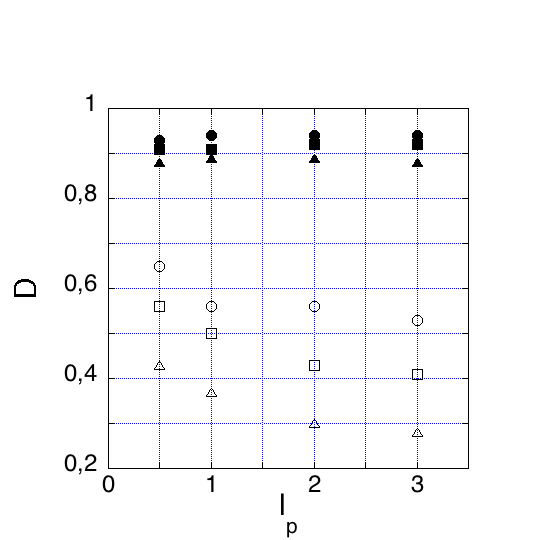}
}
\caption{Upper panel: Diffusion coefficient 
as a function of chain length $N$
for different occupation volumes.
Solid symbols: smallest particle. Blank symbols: largest
 particle.
 Circles denote 6.4\% occupation volume, 
 squares  8\% occupation volume, 
 and triangles 10\% occupation volume.
Lower panel: Diffusion coefficient 
as a function of persistence length $l_p$ for 
different occupation volumes.
 Circles: 46 chains with $N=1000$ 
 (4.6\% occupation volume), 
 squares: 46 chains with $N=1331$ 
 (6.12\% occupation volume), 
 triangles: 46 chains with $N=1728$ 
 (7.95\% occupation volume).
\label{fig:gamma}}
\end{figure}

A variation of the chain length of sufficiently long chromatin 
fibers ($N\in\{50,250,500\}$), while keeping the occupation
volume constant, did not affect the 
diffusion coefficient of the walker. However, once shorter fibers
were involved, with $N<50$, a reduction of $D$ was observed (Fig. \ref{fig:gamma}, 
upper panel). Here
we are approaching the limit of disconnected obstacles ($N = 1$)
which was discussed above.

Next, simulations were carried out with 46
long fibers of length $N \in\{10^3,11^3,12^3\}$,
similar to chromatin fiber territories inside the cell nucleus
(Fig.\  \ref{fig: conformation}). Here, the persistence length
$l_p$ was modified in order to study how particle diffusion
is influenced by local variations of the fiber conformations.
It was found that the diffusion of the smallest 
particle was not  affected by the persistence length
(Fig. \ref{fig:gamma}, lower panel). A larger random walker, however,
was slowing down with increasing persistence length. \\

Furthermore, we observed that the effective volume of the
fiber network changed with different persistence lengths.
After rescaling with respect to the effective volume fraction
of the chains, the data, diffusion coefficients dependent
on different $l_p$, fell on the master curve 
(Fig.\ \ref{fig:geoeff}, lower panel, crosses, stars). This implies that there are
many parameters influencing the particle diffusion such
as chain density, conformation and persistence length,
but these parameters can be reduced to a single one,
the effective volume.

\section{Summary}
It is a well known, if not trivial, wisdom that the complexity
of biological systems requires drastic approximations in order to
become feasible for numerical simulations. The question is: Which
approximations?

The present work was intended to shed some light on the numerical
modelling of diffusion processes inside the cell nucleus. First
of all we have demonstrated that such a process can be described
with a lattice model of the nucleus, the fibers and the random
walk of the protein. The validity of this approach was checked using
a straight comparison with the corresponding continuum model.
Even though the treatment of a living system via lattice model may
impose a cultural shock to some biologists, the benefits of having
a finite number of states and, with the BFM, a highly 
efficient technique to simulate chains at any density, deliver
a substantial speed-up without any apparent loss of accuracy.   

To create a crowded environment of chromatin fibers in
the cell nucleus, three different models were tested, 
yielding different structural properties of the chains.
One of them was made of randomly distributed and disconnected
obstacles, the second one a RW chain (in discrete and continuous 
space) and finally the rather realistic SAW chain model with
excluded volume and persistence length.

In several systematic simulations, including walkers of various
sizes, it was shown that conformational variations of the
crowded environment led to visible modifications of the
diffusion coefficients. First of all, and trivially, a larger
occupation volume was slowing down the diffusion of all walkers.
Moreover, it was found that, the more homogeneous
the obstacles were distributed, the lower were the observed
diffusion coefficients, in particular if the walker was of
large size. Non-interconnected obstacles are more
homogeneous than SAW chains, which in turn are more
homogeneous than RW chains.

These effects could be accounted for with the 
concept of effective occupation volume of the system: This
quantity is a convolution of geometric chain volume 
with the walker's size, i.e.\ the effective free space
available to the walker. Different chain models create 
different distributions of the pores (which form the free space
available to the walker), even if the total chain volume
remains unaffected. A modification of the walker size then 
effectively selects the available set of percolation paths 
through the dense network and hence its diffusion speed. 
Once the diffusion coefficient was plotted as a function of 
effective volume, the data fell onto a master curve, 
i.e.\ the  effect of conformational variations on the diffusion 
could be eliminated. Towards the high end of effective chain 
volume, around 75\%, the onset of subdiffusion was 
observed in all models (Fig. \ref{fig: msd}). The percolation threshold in case of bondless occupation of sites for a three-dimensional cubic lattice is well known, $p_c=0.31$ (see for instance Strenski et al. \cite{percolation}). Thus, in our system, subdiffusion is visible when the occupation is slightly above the percolation threshold. We suspect that in a much larger system without finite size effects, and for much longer simulation times, we could also observe subdiffusion for occupation volumes closer to the percolation threshold.

In a similar manner, the persistence length of SAW chains
did influence the chain distribution, which exhibited
rather strong density fluctuations in case of 
highly flexible chains.
Again, a scaling with the effective volume of
the system allowed to eliminate these conformational
variations on the diffusion behavior, and the data were
falling onto the same master curve.

As a side aspect we observed that the translational diffusion of
polymer chains in the investigated volume range was 
anomalous as a result of boundary restrictions of the cell. 
Once  the space became practically unlimited, 
the translational diffusion coefficient decreased roughly
inversely proportional to the chain length, as is to be
expected when considering Rouse diffusion. 
It is not yet clear to which extent the experimentally
observed slowing of DNA diffusion was caused by binding 
or by crowding (entanglement) effects \cite{slowdown}.
We believe that the model presented here is feasible to
clarify these questions with the help of further systematic 
simulations.

 \section{Acknowledgement}  
 A.W. thanks H.M. and C.W. for their hospitality during a 
 research stay at the Department of Physics at the Xiamen
 University. A.W. was supported by a scholarship from the
 International PhD program of the German Cancer Research
 Center.


\end{document}